\documentclass[pra]{revtex4}
\usepackage{amssymb}
\usepackage{amsmath}

\newcommand{\vc}[1]{\boldsymbol{#1}}

\begin{document}

\title{Gauge Covariance and Spin Current Conservation in the Gauge Field
Formulation of Systems with Spin-Orbit Coupling}

\author{M. S. Shikakhwa}
\affiliation{Physics Program, Middle East Technical University Northern Cyprus Campus,\\
Kalkanl\i, G\"{u}zelyurt, via Mersin 10, Turkey}
\author{S. Turgut}
\author{N. K. Pak}
\affiliation{Department of Physics, Middle East Technical University,\\
TR-06800, Ankara, Turkey}

\begin{abstract}
The question of gauge-covariance in the non-Abelian gauge-field
formulation of two space-dimensional systems with spin-orbit
coupling relevant to spintronics is investigated. Although, these
are generally gauge-fixed models, it is found that for the class
of gauge fields that are space-time independent and satisfy a U(1)
algebra, thus having a vanishing field strength, there is a
residual gauge freedom in the Hamiltonian. The gauge
transformations assume the form of a space-dependent rotation of
the transformed wave functions with rotation angles and axes
determined by the specific form of the gauge-field, i.e., the
spin-orbit coupling. The fields can be gauged away, reducing the
Hamiltonian to one which is isospectral to the free-particle
Hamiltonian, and giving rise to the phenomenon of persistent spin
helix reported first by B.~A.~Bernevig \emph{et al.}
[Phys.~Rev.~Lett. \textbf{97}, 236601 (2006)]. The investigation
of the global gauge transformations leads to the derivation of a
continuity equation where the component of the spin-density along
given directions, again fixed by the specific form of the gauge
field, is conserved.
\end{abstract}

\maketitle

\section{Introduction}

Spintronics\cite{spintronics} is an emerging direction in solid
state physics that is rapidly growing. An important ingredient in
spintronics is the issue of the generation and manipulation of
spin current. An important class of systems where this can be
realized is the quasi-two-dimensional electron (or hole) systems,
where the generation of the spin current might be achieved via the
spin-orbit coupling mechanism\cite{wrinkler}. The most popular
models with various spin-orbit forms are the Rashba
(R)\cite{rashba} and Dresselhaus (D)\cite{dresselhaus} couplings
or a combination of these (R-D). The spin current, its definition
and conservation received considerable attention in the past. Many
works are devoted to the investigation of the question of what is
the correct (or, more precisely, the most convenient) expression
of the spin current that one should use. This is because, unlike
the charge (or matter) current, the spin current is not conserved
in the presence of a magnetic field and/or spin-orbit coupling,
and the continuity equation contains a nonzero right-hand side:
\[
  \frac{\partial S^{a}}{\partial t}+\vc{\nabla}\cdot\vc{J}_{0}^{a}\neq0
\]
where
\begin{equation}
  S^{a}=\phi^{\dagger}\sigma^{a}\phi~,\qquad
  \vc{J}_{0}^{a}=\frac{\hbar}{2mi}\phi^{\dagger}\sigma^{a}\vc{\nabla}\phi +  \textrm{c.c.}
  \label{eq:spin density and bare current}
\end{equation}
are the spin density and the spin current respectively. This led
to works suggesting alternative definitions of the spin
current\cite{shen&xie05,shi.et.al06,relativistic06,Li&Tao07,classical-current08}
that are conserved, other than the one given in Eq.~(\ref{eq:spin
density and bare current}), which is sometimes called the
``natural definition'' and merely generalizes the probability
current density by ``plugging'' in a Pauli matrix in its
definition.

An approach for investigating various aspects of models of
non-relativistic spin one-half particles with spin-orbit coupling,
based on expressing the spin-orbit coupling as an $SU(2)$ gauge
field was introduced in Ref.~\onlinecite{Jin.et.al-JPA} building
on the idea introduced in Ref.~\onlinecite{frohlich93}, and
generated some
interest\cite{Tokatly08,Dartora08,chen&chang08,yang.et.al08,medina-epl08}.
In these models ---in contrast to particle physics models--- the
gauge field, being just the electric field, is physical and
directly observable. Therefore, a gauge transformation corresponds
to switching from a given physical configuration to another, which
is physically different. In the works in
Refs.~\onlinecite{Jin.et.al-JPA,frohlich93,Tokatly08}, a term
quadratic in the gauge field, which ---generally speaking---
breaks the gauge-symmetry of the Hamiltonian, was absorbed into
the definition of the electrostatic potential present in the
Hamiltonian, and thus the Hamiltonian analyzed was taken as
gauge-symmetric. For spin-orbit couplings derived as
$O\left(\frac{1}{c^{2}}\right)$ limit of the Dirac Lagrangian,
this gauge symmetry is, therefore, an
$O\left(\frac{1}{c^{2}}\right)$ symmetry. For spin-orbit couplings
in quasi-two-dimensional electron systems, such as R and D
spin-orbit couplings mentioned above, the origin of these
couplings are different, and it might not be well-justified to
ignore this gauge symmetry-breaking term. For example, in
Refs.~\onlinecite{chen&chang08,yang.et.al08,medina-epl08}, this
gauge-symmetry breaking term was kept. However, the issue of
$SU(2)$ gauge transformations and gauge-symmetry was not fully
analyzed in any of these works. It is the aim of this work to
address this point and investigate some of its consequences. We
carry out the analysis within a Hamiltonian rather than a
Lagrangian framework, which is, we believe, more convenient for
condensed-matter applications.

We also note that while the natural framework for the derivation
of the conserved spin-current is Noether's theorem, there is no
published work that applies this theorem to the free Pauli
Lagrangian to extract the conserved current. We do this in the
appendix.

\section{Spin-Orbit Coupling and SU(2) Gauge Symmetry}

\subsection{SU(2) gauge Symmetry as a Unitary Transformation of the Hamiltonian}
\label{sec:GaugeIsSimilarityTr}
Let the Hamiltonian $H$ (time-independent) of a physical system having eigenstates
$\phi_{n}$ transforms under some unitary transformation $U$ as
$H\rightarrow{}H'=UHU^{-1}$, then $\phi_{n}$ transform as
$\phi_{n}\rightarrow\phi_{n}'=U\phi_{n}$, so that $\phi'_{n}$ are
now eigenstates of $H'$ with the same eigenvalues. In the case of
Abelian gauge transformations, the Hamiltonian is a function of
some gauge field $\vc{A}$,
\begin{equation}
  H=H(\vc{A})=\frac{1}{2m}\left(\vc{p}-\frac{e}{c}\vc{A}\right)^{2} + V(\vc{x})~~.
\end{equation}
A gauge transformation is defined as a simultaneous transformation
of the wavefunction
$\phi_{n}\rightarrow\phi'_{n}=U\phi_{n}=\exp(\frac{-ie}{\hbar
c}\Lambda(\vc{x}))\phi_{n}$ and the gauge field
$\vc{A}\rightarrow\vc{A}'=\vc{A}+\vc{\nabla}\Lambda$. The
transformation of the gauge field so defined guarantees that
$H'=H(\vc{A}')=UH(\vc{A})U^{-1}$, so that $\phi'_{n}=U\phi_{n}$ is
an eigenstate of $H'$. In the non-Abelian $SU(2)$ case, the
Hamiltonian is a function of the non-Abelian gauge field $\vc{W}$
with components $W_{i}=W_{i}^{a}\tau_{a}$, where $\tau_{a}$ are
$2\times2$ matrices satisfying the algebra
$\left[\tau_{a},\tau_{b}\right]=i\epsilon_{abc}\tau_{c}$
($a,b,c=1,2,3$) (e.g., $\tau_a=\sigma_a/2$ where
$\sigma_a$ are Pauli spin matrices),
\begin{equation}
   H=H(\vc{W})=\frac{1}{2m}\left(\vc{p}-g\vc{W}\right)^{2}+V(\vc{x})~~.
   \label{eq:H}
\end{equation}
The spinor wavefunction transforms as
$\phi_{n}\rightarrow\phi'_{n}=U\phi_{n}=\exp(-i\vc{\Lambda}(\vc{x})\cdot\vc{\tau})\phi_{n}$,
and the gauge field transforms simultaneously
$\vc{W}\rightarrow\vc{W}'$. The transformation law for the gauge
field can be derived by demanding that it should ---as in the
Abelian case--- correspond to a transformation of the Hamiltonian
$H(\vc{W'})=UH(\vc{W})U^{-1}$. Since
\begin{align}
   UH(\vc{W})U^{-1}
      &=\frac{1}{2m}U\left(\vc{p}-g\vc{W}\right)U^{-1}\cdot U\left(\vc{p}-g\vc{W}\right)U^{-1}+V(\vc{x})  \\
      &= \frac{1}{2m}\left(\vc{p}-g\vc{W'}\right)\cdot\left(\vc{p}-g\vc{W}'\right)+V(\vc{x})~~,
\end{align}
it can be seen that we need
$U\left(\vc{p}-g\vc{W}\right)U^{-1}=\left(\vc{p}-g\vc{W}'\right)$,
which gives the following transformation law for $\vc{W}$:
\begin{equation}
   \vc{W}'=U\vc{W}U^{-1}+\frac{i\hbar}{g}U\vc{\nabla}U^{-1}
   \label{eq:gauge-trans}
\end{equation}

This is the well-known transformation law for a non-Abelian gauge
field, usually derived within a Lagrangian formalism within the
framework of relativistic field theory models as a transformation
that leaves the action functional invariant\cite{Weinberg}. The
above ``Hamiltonian'' derivation, on the other hand, ascribes a
different meaning to gauge-transformation: A Hamiltonian
$H(\vc{W})$ that transforms under Eq.~(\ref{eq:gauge-trans}) to
another Hamiltonian $H(\vc{W'})$ is unitarily-equivalent to the
original one by $H(\vc{W'})=UH(\vc{W})U^{-1}$, in which case the
Schr\"{o}dinger equation is gauge-covariant. We will also use the term
gauge-covariant or gauge-symmetric for a Hamiltonian that after a
gauge transformation transforms to a form that is unitarily
equivalent to its form before gauge transformation. The
Hamiltonian, Eq.~(\ref{eq:H}), is gauge-covariant by construction.

\subsection{SU(2) Gauge Field Formalism and Gauge Transformations}

The spin-orbit interaction emerges upon considering the
$O\left(\frac{1}{c^{2}}\right)$ expansion of the Dirac Hamiltonian
for a spin one-half particle subject to a scalar potential
$V\left(\vc{x}\right)$:
\[
  H=\frac{\vc{p}^{2}}{2m}+V\left(\vc{x}\right)+\frac{e\hbar}{4m^2c^{2}}\vc{\sigma}\cdot\left(\vc{E}\wedge\vc{p}\right)
   =\frac{\vc{p}^{2}}{2m}+V\left(\vc{x}\right)+\frac{e\hbar}{4m^2c^{2}}\vc{p}\cdot\left(\vc{\sigma}\wedge\vc{E}\right)
\]
where $\vc{\nabla}\wedge\vc{E}=0$ was assumed. Let us now define
the $SU(2)$ gauge field $W_{i}^{a}$ by
\begin{equation}
  -gW_{i}^{a}  \equiv  \frac{e\hbar}{2mc^{2}}\epsilon_{iaj}E_{j} ~~.
  \label{eq:W_as_a_function_of_E_field}
\end{equation}
Using $\vc{W}$, the Hamiltonian can be expressed as
\begin{equation}
  H=\frac{\vc{p}^{2}}{2m}-\frac{g}{m}\vc{p}\cdot\vc{W}+V(\vc{x})=\frac{\vc{p}^{2}}{2m}-\frac{g}{m}\vc{W}\cdot\vc{p}+V(\vc{x})\quad.
\end{equation}
Completing the square, we can put this into the form
\begin{equation}
  H=H(\vc{W})=\frac{\left(\vc{p}-g\vc{W}\right)^{2}}{2m}-\frac{g^{2}}{2m}\vc{W}\cdot\vc{W}+V\left(\vc{x}\right)
  \label{eq:HW}
\end{equation}
where $W_i=W_i^a\tau_a$ is the $i$th component of the field
$\vc{W}$. The above is the Hamiltonian of a spin one-half particle
coupled to the $SU(2)$ gauge field $\vc{W}$. In most of the works
that use this model, the term quadratic in the gauge field, i.e.,
$-\frac{g^{2}}{2m}\vc{W}\cdot\vc{W}$, which generally breaks
the $SU(2)$ gauge symmetry, is absorbed into the potential
$V\left(\vc{x}\right)$ and so, in a sense, it is ``put under the
carpet'', so that the Hamiltonian becomes gauge-invariant
\cite{Jin.et.al-JPA,Tokatly08}. This work, on the other hand, is based on
keeping this term explicit in the Hamiltonian, and analyzing the
model accordingly. In fact, important consequences of keeping this
term explicitly in the Hamiltonian were noted in
Ref.~\onlinecite{medina-epl08}.

Many of the popular models for a two-dimensional electron gas with
spin-orbit couplings, like the Rashba (R) coupling\cite{rashba},
\begin{equation}
  H=\frac{\vc{p}^{2}}{2m}+\frac{\alpha}{\hbar}\left(p_{y}\sigma_{x}-p_{x}\sigma_{y}\right)+V(\vc{x})~~,
  \label{eq:R}
\end{equation}
the Dresselhaus (D) coupling\cite{dresselhaus}
\begin{equation}
  H=\frac{\vc{p}^{2}}{2m}+\frac{\beta}{\hbar}\left(p_{x}\sigma_{x}-p_{y}\sigma_{y}\right)+V(\vc{x})
\end{equation}
and the R-D coupling
\begin{equation}
  H=\frac{\vc{p}^{2}}{2m}
           +\frac{\alpha}{\hbar}\left(p_{y}\sigma_{x}-p_{x}\sigma_{y}\right)
           +\frac{\beta}{\hbar}\left(p_{x}\sigma_{x}-p_{y}\sigma_{y}\right) +V(\vc{x})
  \label{eq:RD}
\end{equation}
can be cast in the form in Eq.~\eqref{eq:HW} \emph{with the term
quadratic in the gauge field present}. Note that, these are
\emph{effective Hamiltonians} that describe the motion of
electrons inside a two-dimensional semiconductor heterostructure.
The spin-orbit coupling terms given above originate from the
strong electric fields associated with the confining potentials of
these structures and the molecular potentials, which are rapidly
varying on the microscopic scale. However, various approximations
(e.g., the envelope wave function approximation) are usually
employed\cite{Bastard} for eliminating the degree of freedom
perpendicular to the 2D confinement plane and averaging out the
molecular potentials on the microscopic scale;  the effective
Hamiltonians given in Eqs.~(\ref{eq:R}-\ref{eq:RD}) are obtained
after such approximations. In these Hamiltonians, $m$ is the
effective mass and $V(\vc{x})$ is an external potential that may
be applied on the structure, which is always slowly varying on
microscopic length scales. For this reason, even though the
spin-orbit terms originate from a physical electric field, these
fields do not appear elsewhere in the effective Hamiltonian. Thus,
we consider the gauge field $\vc{W}$ in these models to be
arbitrary and independent of the external potential $V(\vc{x})$ as
it may have come from any origin. Because of the same reason, the
quadratic term in $\vc{W}$ is not usually negligible in these
effective Hamiltonians.

We will study the motion in 2D and therefore we require $\vc{W}$ to have
$x$ and $y$-components only, and to satisfy the Coulomb gauge-fixing
condition
\begin{equation}
  \partial_{i}W_{i}=0\label{eq:coulomb}~~.
\end{equation}
If the spin-orbit coupling is related to a physical electric field
$\vc{E}$ by Eq.~(\ref{eq:W_as_a_function_of_E_field}), then the
Coulomb condition is equivalent to the Maxwell equation
$\vc{\nabla}\wedge\vc{E}=0$; the spin-orbit gauge theory is a
gauge-fixed theory\cite{N-A-Hydrodynamics07}. While, as we have
mentioned above, we consider the gauge field in our model to have
come from any origin, we are going to assume that the Coulomb
condition is satisfied for such cases. However, it is possible in
some cases to find residual gauge transformations that respect
this gauge condition. A closely related question in relativistic
non-Abelian gauge theories is the Gribov ambiguity\cite{gribov}.

In this subsection, we are going to derive the class of gauge
transformations that respect the Coulomb gauge-fixing condition,
Eq.~(\ref{eq:coulomb}), and under which the Hamiltonian,
Eq.~(\ref{eq:HW}), is gauge-covariant (in the sense discussed in
the last paragraph of Sec.~\ref{sec:GaugeIsSimilarityTr}), and investigate the
consequences and the physical meaning of these transformations.
Even though, the Hamiltonian given in Eq.~(\ref{eq:HW}) will be
our main concern in this article, for the sake of completeness,
the rules will be derived for the general time-dependent gauge transformations.
In the general case, the electron may also be
subjected to (time-dependent) electromagnetic fields. For this
reason, consider the Hamiltonian
\begin{align}
  H &=\frac{1}{2m}\left(\vc{p}-\frac{e}{c}\vc{A}\right)^2-\frac{g}{m}\vc{W}\cdot\left(\vc{p}-\frac{e}{c}\vc{A}\right)
              +V(\vc{x},t) +gW_0 \\
    &=\frac{1}{2m}\left(\vc{p}-\frac{e}{c}\vc{A}-g\vc{W}\right)^2-\frac{g^2}{2m}\vc{W}\cdot\vc{W}+V(\vc{x},t) +gW_0
    \label{eq:Hamiltonian_full}
\end{align}
where $W_0=W_0^a\tau_a$ is the scalar counterpart of gauge field
$\vc{W}$. If $H$ is the original Hamiltonian written for a given
heterostructure (in other words, no SU(2) gauge transformation has
been applied yet), then this term is the same as the Zeeman term,
i.e., we have $g W_0^a = -2\mu_B B^a$ where
$\vc{B}=\vc{\nabla}\wedge\vc{A}$ is the magnetic field and $\mu_B$
is the Bohr magneton. Of course, this is a gauge specific
relation. Such a simple relation between $W_0$ and the actual
magnetic field $\vc{B}$ will not hold when an SU(2) gauge
transformation has already been carried out. The comments that we
have made for the $\vc{W}$ field can be repeated for the scalar
component $W_0$. Just like the SU(2) gauge transformation alters
the ``physical electric field'' that gave way to the spin-orbit
coupling, the transformation also alters the ``physical magnetic
field'' which is associated with the $W_0$ term.The electromagnetic
potentials $V$ and $\vc{A}$ remain invariantunder these transformations.
For this reason, in order to be able to study the SU(2) gauge transformations for the Hamiltonians of
the form given in Eq.~\eqref{eq:Hamiltonian_full}, it is necessary
to take $W_0$ and $\vc{A}$ to be independent. Hence, no specific
relation between $\vc{A}$ and $W_0^a$ should be assumed.

It is known that the Hamiltonian in
Eq.~\eqref{eq:Hamiltonian_full} has the U(1)
symmetry\cite{Jin.et.al-JPA}. We only need to analyze its SU(2)
symmetry. Using the covariance of the Schr\"{o}dinger's equation
$i\hbar\partial_t\psi=H\psi$, the gauge transformation
$\psi^\prime = U\psi$ implies that the Hamiltonian changes as
$H^\prime=UHU^{-1}-i\hbar U\partial_t U^{-1}$ and hence, if the
Hamiltonian remains covariant, the following relations must be
satisfied
\begin{align}
  \vc{W}^\prime   &=   U\vc{W}U^{-1}+\frac{i\hbar}{g}U\vc{\nabla}U^{-1}~~,  \\
  W_0^\prime      &=   UW_0U^{-1}-\frac{i\hbar}{g}U \partial_t U^{-1}~~,  \\
  \vc{W}^\prime\cdot\vc{W}^\prime      &= U \vc{W}\cdot \vc{W}U^{-1}~~.
\end{align}
Obviously, only the quadratic term of the Hamiltonian, namely
$-\frac{g^{2}}{2m^{2}}\vc{W}\cdot\vc{W}$, breaks the
full gauge symmetry of the Hamiltonian. If $H$ is to be gauge-covariant under
the gauge transformation, we should have
\begin{equation}
  g^{2}\vc{W'}\cdot\vc{W'} =\left( \hbar^{2}\left(\vc{\nabla}U\right)\cdot\vc{\nabla}U^{-1}+
  i\hbar g\Big[ U\vc{W}\cdot \vc{\nabla}U^{-1}-\left(\vc{\nabla}U\right)\cdot\vc{W}U^{-1}\Big]\right)+g^{2}U\vc{W}\cdot\vc{W}U^{-1}~~.
\end{equation}
Gauge-covariance then, dictates that the first bracket above should
vanish
\begin{equation}
  \hbar^{2}\left(\vc{\nabla}U\right)\cdot\vc{\nabla}U^{-1} +
  i\hbar g\Big[ U\vc{W}\cdot \vc{\nabla}U^{-1}-\left(\vc{\nabla}U\right)\cdot\vc{W}U^{-1}\Big]=0~~.
  \label{eq:quadra-condition}
\end{equation}
Similarly, forcing $\vc{W}'$ to respect the Coulomb gauge
condition we have
\begin{equation}
   i\hbar g\vc{\nabla}\cdot\vc{W}' = -\hbar^2 \Big\{\left(\vc{\nabla}U\right) \cdot \vc{\nabla}U^{-1} +  U\nabla^{2}U^{-1}\Big\}+
   i\hbar g\Big[ U\vc{W}\cdot \vc{\nabla}U^{-1} + \left(\vc{\nabla}U\right)\cdot\vc{W}U^{-1}\Big]=0~~.
   \label{eq:coulomb-condition}
\end{equation}
Adding the above two equations leads to
\begin{equation}
  \left(\vc{\nabla}-\frac{2ig}{\hbar}\vc{W}\right)\cdot\vc{\nabla}U^{-1}=0~~.
  \label{eq:covariance condition}
\end{equation}
What is remarkable is that the Eq.~\eqref{eq:covariance condition}
is in fact equivalent to the previous two conditions, namely
Eqs.~\eqref{eq:quadra-condition} and \eqref{eq:coulomb-condition}.
This can be shown by inserting into
Eqs.~\eqref{eq:quadra-condition} and \eqref{eq:coulomb-condition}
the following expressions
\begin{align*}
  ig U  \vc{W}\cdot\vc{\nabla}U^{-1}&=\frac{\hbar}{2}U\nabla^2 U^{-1}~~, \\
  ig \vc{\nabla}U  \cdot \vc{W}U^{-1}&=-\frac{\hbar}{2}(\nabla^2U) U^{-1}~~, \\
\end{align*}
where the former is a re-expression of Eq.~\eqref{eq:covariance
condition} and the latter is obtained from the hermitian
conjugation of the former. Hence, Eq.~\eqref{eq:covariance
condition} is the differential equation that gives the class of
gauge transformations $U$ that respect the Coulomb's gauge
condition, and under which the Hamiltonian is gauge-covariant. The
equation is valid for time-dependent transformations and in
the presence of electromagnetic fields. However, for the sake of
simplicity, in the rest of the article the Hamiltonian is taken to
be as in Eq.~\eqref{eq:HW} (i.e., $\vc{A}=0$ and $H$ is time
independent) and only time-independent transformations are
discussed.

Rather than attempting to find all of the solutions of
Eq.~\eqref{eq:covariance condition} systematically, which is an
involved task, we start by noting that a class of solutions of the
above equation that one can easily \emph{guess} is
\begin{equation}
  U=\exp\left(-\frac{2ig}{\hbar}\vc{W}\cdot\vc{x}\right)
  \label{eq:explicit U}
\end{equation}
valid for $\vc{W}$ satisfying the following conditions:
\begin{eqnarray}
  \left[W_1,W_2\right] &=& 0~~,\label{eq:commutator i,j} \\
  x_{j}\partial_{i}W_{j} &=& 0~~(i=1,2)~. \label{eq:2nd condition}
\end{eqnarray}
The first condition above means that the field $\vc{W}$ is now
Abelian, i.e., $U(1)$, although it is still a $2\times2$ matrix. It
is straightforward to check that the above $U^{-1}$ satisfies
both conditions (\ref{eq:quadra-condition}) and
(\ref{eq:coulomb-condition}). We now move on to try to find explicitly
various fields $\vc{W}$ satisfying the above two conditions so as
to identify spin-orbit couplings that lead to gauge-covariant
Hamiltonians.

The condition in Eq.~(\ref{eq:2nd condition}) is satisfied trivially by
restricting our fields to those that are space-time independent.
This amounts then to considering fields that lead to a vanishing
field strength tensor, i.e.,
\begin{equation}
F_{ij}=\partial_{i}W_{j}-\partial_{j}W_{i}-\frac{ig}{\hbar}\left[W_{i},W_{j}\right]=0
\end{equation}
To this end, we first analyze condition (\ref{eq:commutator i,j})
more closely. Writing $W_{i}=W_{i}^{a}\tau_{a}$, and taking $a$
to run over 1 and 2 only (i.e., assuming that $W_{i}^{3}=0$), this condition can be expressed as
\begin{equation}
  \left[W_{1},W_{2}\right]=i\tau_{3}\left(W_{1}^{1}W_{2}^{2}-W_{1}^{2}W_{2}^{1}\right)=0~~,
  \label{eq:commutator 1,2}
\end{equation}
which can be also written in terms of a determinant
\begin{equation}
  \left\vert \begin{array}{cc}
  W_{1}^{1} & W_{1}^{2}\\
  W_{2}^{1} & W_{2}^{2}
  \end{array}\right\vert=0~~.
  \label{eq:determinant}
\end{equation}
So, our task reduces to the trivial task of finding sets of
$2\times2$ constant matrices with vanishing determinants. The
possibilities are obviously infinite! We list a few of these in
Table \ref{Table:zerodet}. $\alpha$ and $\beta$ are constants, and the notation is
self-explanatory.

\begin{table}
\caption{Some special spin-orbit couplings that has vanishing determinant $\det (W_i^a)=0$.}
\begin{tabular}{|l|l|}
\hline
~$W_{i}^{a}=\left(\begin{array}{cc}
-\alpha & \alpha\\
-\alpha & \alpha\end{array}\right)$ & ~$\vc{W}=\left(-\alpha\tau_{x}+\alpha\tau_{y},-\alpha\tau_{x}+\alpha\tau_{y}\right)$ \\
\hline
~$W_{i}^{a}=\left(\begin{array}{cc}
\alpha & \alpha\\
-\alpha & -\alpha\end{array}\right)$ & ~$\vc{W}=\left(\alpha\tau_{x}+\alpha\tau_{y},-\alpha\tau_{x}-\alpha\tau_{y}\right)$\\
\hline
~$W_{i}^{a}=\left(\begin{array}{cc}
\alpha & \beta\\
\alpha & \beta\end{array}\right)$ & ~$\vc{W}=\left(\alpha\tau_{x}+\beta\tau_{y},\alpha\tau_{x}+\beta\tau_{y}\right)$\\
\hline
~$W_{i}^{a}=\left(\begin{array}{cc}
\alpha & \alpha\\
\beta & \beta\end{array}\right)$ & ~$\vc{W}=\left(\alpha\tau_{x}+\alpha\tau_{y},\beta\tau_{x}+\beta\tau_{y}\right)$\\
\hline
\end{tabular}
\label{Table:zerodet}
\end{table}

The first two fields in Table \ref{Table:zerodet} are of special
importance as they represent the R-D spin-orbit coupling
with constant coupling coefficients in the special cases of
$\alpha=\pm\beta$, respectively, a model that was studied
extensively in the literature. Here, they appear just as two
members of an infinite set of possibilities. Other field
configurations have no physical realizations as far as we know.

To gain a deeper insight on the meaning of the gauge transformation,
Eq.~(\ref{eq:explicit U}), we first investigate its effect on the
gauge field itself $\vc{W}$;
\begin{equation}
  \vc{W}\rightarrow\vc{W}'=\frac{i\hbar}{g}U\vc{\nabla}U^{-1}+U\vc{W}U^{-1}=-2U\vc{W}U^{-1}+U\vc{W}U^{-1}=-U\vc{W}U^{-1}=-\vc{W}
\end{equation}
Our gauge transformation, under which the Hamiltonian is gauge-covariant
amounts merely to reversing the direction of the gauge field, i.e.,
the electric field generating the spin-orbit coupling, which is a
satisfying result. To investigate the effect of the corresponding phase
transformation on the wave function, we note that Eq.~(\ref{eq:explicit U})
can be expressed as
\begin{equation}
  U = \exp\left(-i\vc{\eta}\cdot\vc{\tau}\right) = I\cos\frac{\eta}{2}-i\vc{\hat{n}}\cdot\vc{\sigma}\sin\frac{\eta}{2}
  \label{eq:local rotation}
\end{equation}
where
\begin{equation}
  \eta^{a}=\frac{2g}{\hbar}W_{i}^{a}x^{i}
\end{equation}
with both $a$ and $i$ running over 1 and 2 only, $\eta=\vert\vc{\eta}\vert$ and $\vc{\hat{n}}=\vc{\eta}/\eta$. The gauge
transformation $U$ is, therefore, a rotation about the axis
$\vc{\hat{n}}$ with a \emph{space-dependent} angle $\eta$. To consider
a specific example, consider $W_{i}^{a}=\left(\begin{array}{cc}\alpha & \beta\\ \alpha & \beta \end{array}\right)$
(or  $\vc{W}=(\alpha\tau_{x}+\beta\tau_{y},\alpha\tau_{x}+\beta\tau_{y})$),
which is the third entry in Table \ref{Table:zerodet} above, for which we have
$\eta=(x+y)\sqrt{\alpha^{2}+\beta^{2}}$ and
$\vc{\hat{n}}=(\alpha\vc{\hat{\imath}}+\beta\vc{\hat{\jmath}})/\sqrt{\alpha^{2}+\beta^{2}}$.
Thus
\begin{equation}
  U=I\cos\left((x+y)\sqrt{\alpha^{2}+\beta^{2}}\right)-i\vc{\hat{n}}\cdot\vc{\sigma}\sin\left((x+y)\sqrt{\alpha^{2}+\beta^{2}}\right)~~.
\end{equation}
Here, the space-dependence of the rotation angle in the arguments
of the trigonometric functions above is evident.

It is a well-established fact that the vanishing of the field
strength tensor of any gauge-field is a necessary and sufficient
condition for the existence of a gauge transformation that takes
this field to zero\cite{Weinberg}. Therefore, it should be
possible to transform all the gauge fields under consideration to
zero. Finding the transformation that achieves this is an easy
task. One immediately checks that for
\begin{equation}
  U=\exp\left(-\frac{ig}{\hbar}\vc{W}\cdot\vc{x}\right)
  \label{eq:removing field}
\end{equation}
one has:
\begin{equation}
  \vc{W}\rightarrow\vc{W}'=\frac{i\hbar}{g}U\vc{\nabla}U+U\vc{W}U^{-1}=-U\vc{W}U^{-1}+U\vc{W}U^{-1}=0~~.
  \label{eq:gauge-away}
\end{equation}
Obviously, the above transformation can also be brought to the
form of a rotation about some axis with a position-dependent
angle, just as was done with the gauge transformation,
Eq.~(\ref{eq:local rotation}). The work in
Ref.~\onlinecite{exact-su(2)06} considered a specific field; the
one that results from the R-D coupling in the special cases
$\alpha=\pm\beta$ (the first and second entries in Table
\ref{Table:zerodet}). The gauge field was gauged away using a
transformation identical to the one above leaving a free particle
Hamiltonian with the same spectrum as that of the spin-orbit
coupled one. This fact was employed to account for the appearance
of a persistent spin helix (PSH) in this model: One can imagine a
free particle that enters a region where the spin-orbit coupling
is turned on, which corresponds to a gauge transformation that is
the inverse of the one given in Eq.~(\ref{eq:gauge-away}), so the
particle is subject to a position-dependent rotation about some
axis that is similar to the one discussed above. The particle will
propagate with its spin rotating so that its projection along the
rotation axis is conserved, a phenomenon that was called the
PSH\cite{exact-su(2)06}. Here, we are saying that it is possible
to gauge away any space-time independent field satisfying the
condition (\ref{eq:commutator 1,2}). In other words, any
Hamiltonian with a gauge-field satisfying the condition
(\ref{eq:commutator 1,2}) is in fact unitarily equivalent to the
free particle Hamiltonian, and we can have the phenomenon of PSH
in all these cases.

There is a fine but important detail here, which was not noted, or
at least not discussed in the literature. The spin-orbit coupled
Hamiltonian, Eq.~(\ref{eq:HW}) is not in fact covariant under the
transformation, Eq.~(\ref{eq:gauge-away}). The reason is the
quadratic term $g^{2}\vc{W}\cdot\vc{W}$, which is non-zero in one
Hamiltonian and vanishes in the other one. Gauge-covariance, on
the other hand, as we have noted earlier, requires that it
transforms as $g^{2}\vc{W}\cdot\vc{W}\rightarrow
g^{2}U\vc{W}\cdot\vc{W}U^{-1}=g^{2}\vc{W}\cdot\vc{W}$!
Fortunately, for space-time independent fields, this quadratic
term is just a constant. Thus, the gauge-transformed Hamiltonian
is gauge-covariant up to a constant, and is unitarily equivalent
to a free particle Hamiltonian up to a constant;
\begin{equation}
  H\left(\vc{W'}=0\right)=UH\left(\vc{W}\right)U^{-1}+\frac{g^{2}}{2m}\vc{W}\cdot\vc{W}=UH\left(\vc{W}\right)U^{-1}+\mathrm{constant}=H_{0}.
  \label{eq:unitary equivalence}
\end{equation}
In the work in Ref.~\onlinecite{chen&chang08}, where the R-D
spin-orbit coupling in the special case $\alpha=\beta=$constant
was considered, the gauge field was gauged away using exactly the
same transformation above. However, there, the transformation of
the constant quadratic term was ignored altogether. Being just a
constant, the result is the same. However, strictly speaking, the
transformation of this term should be considered. Our treatment
here illuminates this point. It also suggests that there exist
---in principle--- a wide class of spin-orbit coupling forms that
lead to essentially the same result, i.e., the unitary-equivalence
with the free particle Hamiltonian, and consequently the emergence
of the PSH. Only a few of these are realized physically, the
others are ---so far--- theoretical. Yet, it is highly possible
that physical systems with such couplings might be realized in the
near future.

At this point, we can come back to the question of finding
solutions to Eq.~(\ref{eq:covariance condition}) . Indeed, from
Eq.~(\ref{eq:unitary equivalence}), it is obvious that under the
transformation
\begin{equation}
  U_{\vc{W'}}=\exp\left(\frac{ig}{\hbar}\vc{W'}\cdot\vc{x}\right)
  \label{eq:restoring field}
\end{equation}
the free-particle Hamiltonian (for  $\vc{W}'$ independent of space
and time such that $\det(W_i^a{}')=0$) transforms as:
\begin{equation}
 U_{\vc{W'}}H\left(0\right)U^{-1}_{\vc{W'}}=H\left(\vc{W'}\right)+\frac{g^{2}}{2m}\vc{W'}\cdot\vc{W'}.
  \label{eq:restoring HW'}
\end{equation}
Therefore, we can immediately write down the following
transformation:
\begin{equation}
 U_{\vc{W'}}U_{\vc{W}}H\left(\vc{W}\right)U^{-1}_{\vc{W}}U^{-1}_{\vc{W'}}=H\left(\vc{W'}\right)+\frac{g^{2}}{2m}\vc{W'}\cdot\vc{W'}
 -\frac{g^{2}}{2m}\vc{W}\cdot\vc{W}.
  \label{eq:HW to HW'}
\end{equation}
where we have denoted with $ U_{\vc{W}}$ the transformation (\ref{eq:removing field}).If, now,  the condition $\vc{W}\cdot\vc{W}=\vc{W}'\cdot\vc{W}'$ (or, equivalently $W_i^aW_i^a=W_i^a{}'W_i^a{}'$) is satisfied, then the above equation reduces to
\begin{equation}
 U_{\vc{W'}}U_{\vc{W}}H\left(\vc{W}\right)U^{-1}_{\vc{W}}U^{-1}_{\vc{W'}}=H\left(\vc{W'}\right)
\label{eq:HW to HW'invariance}
\end{equation}
which immediately means that the unitary transformation
\begin{equation}
  U=U_{\vc{W'}}U_{\vc{W}}=\exp\left(\frac{ig}{\hbar}\vc{W}'\cdot\vc{x}\right)\exp\left(-\frac{ig}{\hbar}\vc{W}\cdot\vc{x}\right)~~.
 \label{eq:ultimate U}
\end{equation}
induces  a gauge transformation
$\vc{W}\longrightarrow\vc{W}^\prime$ under which the Hamiltonian
is gauge-covariant, i.e.,
\begin{equation}
   \frac{\left(\vc{p}-g\vc{W}'\right)^{2}}{2m}-\frac{g^{2}}{2m^{2}}\vc{W}'\cdot\vc{W}'+V(\vc{x}) =
   U\left( \frac{\left(\vc{p}-g\vc{W}\right)^{2}}{2m}-\frac{g^{2}}{2m^{2}}\vc{W}\cdot\vc{W}+V(\vc{x})\right) U^{-1}~~,
\end{equation}
Therefore, all fields $\vc{W}$ independent of space and time such
that $\det(W_i^a)=0$ with the same value of the product
$\vc{W}\cdot\vc{W}$ are related by gauge transformations of the
form (\ref{eq:ultimate U}) that are symmetries of the Hamiltonian.
The gauge transformation (\ref{eq:explicit U}) found earlier is
just one of these transformations as can be seen easily by
substituting $\vc{W}^\prime \longrightarrow -\vc{W}$ in
(\ref{eq:ultimate U}). Moreover, although laborious, it is
straightforward to show that this $U$ satisfies
Eq.~(\ref{eq:covariance condition}). Thus, infinite classes of
solutions for this equation have been constructed. Physically, the
above results say that different spin-orbit couplings
corresponding to various electric fields configurations are
related ---if they satisfy certain conditions--- by a
gauge-transformation, and their corresponding Hamiltonians are
unitarily equivalent, thus having the same spectrum.

\section{Global Gauge-Symmetry and Conserved Spin Currents}

We turn now to the investigation of the $SU(2)$ global phase
invariance of the Hamiltonian, Eq.~(\ref{eq:HW}). When $\vc{W}$ is
space-time independent satisfying the commutation relations in
Eq.~(\ref{eq:commutator 1,2}), $H$ is invariant under the gauge
transformation
$U=\exp\left(-\frac{2ig}{\hbar}\vc{W}\cdot\vc{x}\right)$, which is
actually an Abelian symmetry as we have noted earlier. If we
replace $\vc{x}$ with a constant vector $\frac{2g}{\hbar}\vc{l}$,
we will have the Hamiltonian invariant under the global  phase
transformation
\begin{equation}
  U=\exp\left(-i\vc{W}\cdot\vc{l}\right)~~.
  \label{eq:global trans}
\end{equation}
Obviously, this is just a rotation in the spin space. To put it in
a more convenient form, we again write
\[
  \vc{W}\cdot\vc{l}=W_{i}^{a}\tau_{a}l^{i}=\xi^{a}\tau_{a}
\]
with
\begin{equation}
   \xi^{a}=W_{i}^{a}l^{i}~, \quad \xi=\vert\vc{\xi}\vert~, \quad \vc{\hat{n}}=\vc{\xi}/\xi
   \label{eq:define xi}
\end{equation}
so that
\begin{equation}
  U=\exp\left(-i\vc{\xi}\cdot\vc{\tau}\right)=\exp\left(-i\xi\vc{\hat{n}}\cdot\frac{\vc{\sigma}}{2}\right)~.
  \label{eq:rotation}
\end{equation}
The invariance of $H$ under the above transformation immediately
implies the conservation of the operator $\vc{\sigma}\cdot\vc{\hat{n}}$
(in the Heisenberg picture), which means that if $\phi$ is the wavefunction, then
$\int\phi^{\dagger}\vc{\sigma}\cdot\vc{\hat{n}}\phi$ is constant in
time. Therefore, we expect the density
$\vc{S}\cdot\vc{\hat{n}}=\phi^{\dagger}\vc{\sigma}\cdot\vc{\hat{n}}\phi$ to
satisfy a continuity equation with a conserved current of the form
\begin{equation}
  \frac{\partial}{\partial t}\left(\vc{S}\cdot\vc{\hat{n}}\right)+\partial_{i}\left(\vc{J}_{i}\cdot\vc{\hat{n}}\right)=0
  \label{eq:continuity n}
\end{equation}
The above relation then means that the spin density
$\vc{S}\cdot\vc{\hat{n}}$ is conserved, and thus a particle polarized
in this direction will not feel a torque and thus will have a long
life time, a property of great value in spintronics
\cite{Hall.app.phys.lett,schleiman-et.al.diode}. Since $\vc{\hat{n}}$
is determined by $W_{i}^{a}$, i.e., by the specific spin-orbit
coupling present, then for different spin-orbit couplings, we will
have different conserved spin densities.

It is interesting to see how the above continuity equation emerges
from the Schr\"{o}dinger equation in the conventional sense. For
the Hamiltonian in Eq.~(\ref{eq:HW}), with $\vc{W}$ being
\emph{any arbitrary} $SU(2)$ gauge field in the Coulomb gauge, one
gets after some algebra:
\begin{equation}
  \frac{\partial S^{a}}{\partial t}+\vc{\nabla\cdot}\vc{J}^{a} = \frac{-ig}{2m} \epsilon^{abc} \left[\phi^{\dagger}\sigma^{b}W_{i}^{c}\left(\partial_{i}\phi\right) - \left(\partial_{i}\phi^{\dagger}\right)W_{i}^{b}\sigma^{c}\phi\right]
  \label{eq:continuity-broken}
\end{equation}
where the current is $\vc{J}^{a}=\vc{J}_{0}^{a}+{\vc{J}^{a}}'$;
$\vc{J}_{0}^{a}$ is the bare spin current defined earlier in
Eq.~(\ref{eq:spin density and bare current}), and $\vc{J}'^{a}$ is
defined as
\begin{equation}
  \vc{J}'^{a}=-\frac{g}{2m}\phi^{\dagger}\vc{W}^{a}\phi~~.
  \label{eq:J'}
\end{equation}
Obviously the spin current is not conserved due to the presence of
spin-orbit coupling, which ---just as in the case of a magnetic
field--- breaks the $SU(2)$ global phase invariance. The above
continuity equation can be put into an alternative form where the
current is covariantly conserved. To do this, we note that the
right-hand side of Eq.~(\ref{eq:continuity-broken}) can be brought
to the form
$-\frac{g}{\hbar}\epsilon^{abc}\vc{W}^{b}\cdot\vc{J}_{\vc{W}}^{c}$
with
\[
   \vc{J}_{\vc{W}}^{c}\equiv-\frac{i\hbar}{2m}(\phi^{\dagger}\sigma^{a}\vc{D}\phi-\vc{(D}\phi)^{\dagger}\sigma^{a}\phi)
\]
where $\vc{D}=\vc{\nabla}-\frac{ig}{\hbar}\vc{W}$ is the covariant
derivative. Therefore, Eq.~(\ref{eq:continuity-broken}) can be
cast into the form:
\begin{equation}
  \frac{\partial}{\partial t}S^{a}[\vc{W}]+\vc{D}^{ac}\cdot\vc{J}_{\vc{W}}^{c}=0
  \label{eq:covariant continuity}
\end{equation}
where
$\vc{D}^{ac}\equiv\delta^{ac}\vc{\nabla}+\frac{g}{\hbar}\epsilon^{abc}\vc{W}^{b}$.
This is the covariant continuity equation, and was written down in Ref.~\onlinecite{Jin.et.al-JPA} directly from the Lagrangian of the theory
rather than from the equations of motion. To derive the continuity
equation, Eq.~(\ref{eq:continuity n}), from the above equation for
the special cases when the fields $\vc{W}$ satisfy the commutation
relations (\ref{eq:commutator i,j}), we multiply both sides of
Eq.~(\ref{eq:covariant continuity}) by $\hat{n}^{a}$ to get
\begin{eqnarray}
  \frac{\partial}{\partial t} \hat{n}^{a}S^{a}[\vc{W}]+\hat{n}^{a}\vc{D}^{ac}\cdot\vc{J}_{\vc{W}}^{c} & = & 0,
  \label{eq:n-covariant continuity}
\end{eqnarray}
We need to show that
$\epsilon^{abc}\hat{n}^{a}W_{i}^{b}J_{\vc{w},i}^{c}=0$
so as to reduce the covariant derivative to the ordinary
derivative. Expressing $n^{a}$ in terms of $\xi$ and using
Eq.~(\ref{eq:define xi}) we get
\[
  \frac{g}{\hbar}\epsilon^{abc}n^{a}W_{i}^{b}J_{\vc{W},i}^{c}=\frac{g}{\hbar}\frac{l_{j}}{\xi}\epsilon^{abc}W_{j}^{a}W_{i}^{b}J_{\vc{W},i}^{c}=0
\]
where we have noted that $\det (W_i^a)=0$ is equivalent to
$\epsilon^{abc}W_{j}^{a}W_{i}^{b}=0$ for $i,j=1,2$. Thus,
Eq.~(\ref{eq:n-covariant continuity}) immediately reduces to the
continuity equation, Eq.~(\ref{eq:continuity n}).

Again, we apply the above results to explicitly find the conserved
spin and current densities for some of the field examples that we
have presented in Table \ref{Table:zerodet}. For $(W_{i}^{a})=\left(\begin{array}{cc}
-\alpha & \alpha\\
-\alpha & \alpha\end{array}\right)$, so that $\vc{\hat{n}}=-\frac{1}{\sqrt{2}}\left(\vc{\hat{\imath}}-\vc{\hat{\jmath}}\right)=\vc{\hat{n}}_{1}$,
we get
\begin{equation}
  \frac{\partial}{\partial t}\left(\vc{S}\cdot\vc{\hat{n}}_{1}\right)+\partial_{i}\left(\vc{J}_{i}\cdot\vc{\hat{n}}_{1}\right)=0
  \label{eq:continuity n1}
\end{equation}

The above result which corresponds to the R-D coupling with $\alpha=\beta$=constant
was reported in Ref.~\onlinecite{schleiman-et.al.diode}, without
the use of gauge-field formalism, however. In our case, we have re-derived
it within the context of a more general theoretical formalism, and
extended it to other couplings. For $(W_{i}^{a})=\left(\begin{array}{cc}
\alpha & \beta\\
\alpha & \beta\end{array}\right)$, we have
$\vc{S}\cdot\vc{\hat{n}}_{2}$ satisfying the above continuity equation,
with
$\vc{\hat{n}}_{2}=(\alpha\vc{\hat{\imath}}+\beta\vc{\hat{\jmath}})/\sqrt{\alpha^{2}+\beta^{2}}$.
As for $(W_{i}^{a})=\left(\begin{array}{cc}
\alpha & \alpha\\
\beta & \beta\end{array}\right)$ we have $\vc{S}\cdot\vc{\hat{n}}_{3}$
conserved with
$\vc{\hat{n}}_{3}=(\vc{\hat{\imath}}+\vc{\hat{\jmath}})/\sqrt{2}$.

We again note that the natural framework for the derivation of the continuity
equation is Noether's theorem. The infinitesimal version of the transformation,
Eq.~(\ref{eq:global trans}), is a global continuous symmetry of the
Lagrangian of the theory, and Noether's theorem demands the existence
of a conserved charge and current related by a continuity equation,
which can be shown to be just the continuity equation, Eq.~(\ref{eq:continuity n}).
We give the details of this derivation as well in the appendix.

Since the free particle Hamiltonian and the spin-orbit-coupled
Hamiltonian with fields of the type given in Table
\ref{Table:zerodet} are related (up to a constant) by the gauge
transformation, Eq.~(\ref{eq:gauge-away}), it should be possible
to get the continuity equations, Eq.~(\ref{eq:continuity n}), for
these classes of fields from the free-particle continuity equation
by a gauge transformation. Obviously, it will be sufficient to
show how one can relate the free-particle continuity equation to
the continuity equation, Eq.~(\ref{eq:covariant continuity}), through a
gauge transformation. For this purpose, we first express both the
bare and the covariant continuity equations in the adjoint
representation, so we define
\begin{equation}
  \widetilde{S}[\vc{W}]\equiv S^{a}[\vc{W]}\tau_{a}~~,
  \qquad
  \widetilde{J}_{i}[\vc{W}]\equiv J_{i\vc{W}}^{a}[\vc{W]}\tau_{a}~~,
\end{equation}
and express the covariant continuity equation,
Eq.~(\ref{eq:covariant continuity}), as
\begin{equation}
  \frac{\partial}{\partial t}\widetilde{S}[\vc{W}]+\widetilde{D}_{i}[\vc{W}]\widetilde{J}_{i}[\vc{W}]=0~~,
  \label{eq:covariant continuity adjoint}
\end{equation}
where
$\widetilde{D}_{i}[\vc{W}]=\partial_{i}-\frac{ig}{\hbar}[W_{i}^{a}\tau_{a},\ldots]$
is the covariant derivative in the adjoint representation, and the
dependence on the gauge field $\vc{W}$ was now shown explicitly.
Under a gauge transformation, $\vc{W}\rightarrow\vc{W}'$, given by
Eq.~(\ref{eq:gauge-trans}), the above equation transforms to
\begin{equation}
  \frac{\partial}{\partial t}\widetilde{S}[\vc{W}']+\widetilde{D}_{i}[\vc{W}']\widetilde{J}_{i}[\vc{W}']=0~~.
  \label{eq:transformed continuity adjoint}
\end{equation}
Now, the free-particle continuity equation can be expressed also in
the adjoint representation as
\begin{equation}
  \frac{\partial}{\partial t}\widetilde{S}[0]+\widetilde{D}_{i}[0]\widetilde{J}_{i}[0]=0
  \label{eq:free continuity adjoint}
\end{equation}
with
$\widetilde{D}_{i}[0]=\partial_{i}-\frac{ig}{\hbar}[0,\ldots]$.
Under the (inverse of) the gauge transformation given by
Eq.~(\ref{eq:gauge-away}), we have
\begin{equation}
  0\rightarrow\vc{W}=\frac{i\hbar}{g}U^{-1}\vc{\nabla}U,
\end{equation}
so that
$\widetilde{D}_{i}[0]=\partial_{i}-\frac{ig}{\hbar}[0,\ldots]\rightarrow\widetilde{D}_{i}[\vc{W}]=\partial_{i}-\frac{ig}{\hbar}[\vc{W},\ldots]$,
and Eq.~(\ref{eq:free continuity adjoint}) transforms immediately
to Eq.~(\ref{eq:covariant continuity adjoint}).

\section{Summary and Conclusions}

The question of gauge-covariance in the two-dimensional
Hamiltonian of a spin one-half particle subject to spin-orbit
coupling formulated in a non-Abelian gauge field language was
considered. Such a Hamiltonian contains a term that is quadratic
in the gauge field, which generally breaks the gauge symmetry.
Moreover, with the gauge field representing a physical quantity,
namely the electric field, the theory is a gauge-fixed one. The
conditions for the existence of residual gauge symmetry in the
Coulomb gauge were investigated, and a condition for its
existence, namely Eq.~(\ref{eq:covariance condition}), was
derived. A class of gauge fields and their corresponding gauge
transformations that satisfy this condition were found. They turn
out to be fields whose components are any $2\times2$ space-time
independent Abelian matrices , and thus are just gauges having a
vanishing field strength tensor. The corresponding gauge
transformation that are symmetries of the Hamiltonian are seen to
correspond to rotations in the spin space of the particle with
space-dependent rotation angles, the specific form of which and of
the rotation axes being determined by the explicit form of the
gauge field . Gauging away the gauge field leads us to see the
phenomenon of persistent spin helix (PSH) discussed in the
literature\cite{exact-su(2)06}. The global version of the
admissible gauge transformations, which also form a symmetry of
the Hamiltonian, lead to a continuity equation for the projection
of the spin density along the rotation axis, which is fixed by the
specific form of the gauge field. The spin along these axes is not
subject to any torque and is, therefore, long-living; a property
that is important in spintronic applications and was reported in
the literature for the specific case of a special case of the R-D
coupling in Ref.~\onlinecite{schleiman-et.al.diode}. Our
re-derivation adds to this ---as well as to the PSH derivation---
in two ways: First, it comes within a general framework, based on
the idea of gauge-covariance of the Hamiltonian of two-dimensional
systems with spin-orbit coupling. Second; the phenomena reported
in the above two references were derived for the specific cases of
R-D coupling with constant equal coefficients only. In our case,
these two cases appear as only two special cases of wider
theoretical possibilities. The R-D couplings are the only
couplings that are realized physically so far.

Finally, an important point that deserves further attention is
the existence of solutions to Eq.~(\ref{eq:covariance condition})
other than the Abelian space-independent ones found here. Indeed,
it would be interesting to find such solutions; this issue is
under current investigation.

\section*{Acknowledgment}

This work was supported by the Campus Research Fund of Middle East
Technical University Northern Cyprus Campus under project BAP-Fen
11.

\appendix

\section*{Appendix}

In this section, Noether's theorem is applied for deriving the
expression for the spin current, Eq.~\eqref{eq:spin density and
bare current}, and establishing the continuity equation,
Eq.~\eqref{eq:continuity n}. For this purpose, we first look at
the case of bare spin current. Consider the case where there is no
magnetic field and no spin-orbit coupling, i.e., the gauge field
is zero. The Pauli Lagrangian density for the electrons is then
\begin{equation}
  \mathcal{L} = \frac{i\hbar}{2}\left(\phi^{\dagger}\dot{\phi}-\dot{\phi}^\dagger\phi\right)
         - \frac{\hbar^{2}}{2m} \vc{\nabla}\phi^{\dagger}\cdot\vc{\nabla}\phi
         -V(\vc{x})\phi^\dagger \phi    \quad,
  \label{eq:PauliLagrangian}
\end{equation}
where $\phi$ is the two-component spinor wave function. This Lagrangian
density has the global SU(2) symmetry, i.e., it is invariant under the
transformation
$$ \phi \rightarrow \exp(-i\vc{\alpha}\cdot\vc{\sigma})\phi
  \approx \phi -i\vc{\alpha}\cdot\vc{\sigma}\phi\quad,
$$
where $\vc{\alpha}$ is an infinitesimal vector that is independent of
position and time. Noether's theorem\cite{Weinberg} then implies the
conservation of current density $\mathcal{J}^\mu$ (i.e., the continuity
equation $\partial_\mu\mathcal{J}^\mu=0$ is satisfied) where
\begin{equation}
\mathcal{J}^\mu =
      \delta\phi^\dagger \frac{\partial\mathcal{L}}{\partial(\partial_\mu\phi^\dagger)}
     +                   \frac{\partial\mathcal{L}}{\partial(\partial_\mu\phi)} \delta\phi\quad.
  \label{eq:NoetherCurrent}
\end{equation}
It is straightforward to check that the time and position components of
the conserved current are given by
\begin{align}
  \mathcal{J}^0 &= \hbar \alpha_a S^a \quad,\\
  \vc{\mathcal{J}} &= \hbar \alpha_a \vc{J}_0^a \quad,
\end{align}
where $S^a$ and $\vc{J}_0^a$ are as given in Eq.~\eqref{eq:spin density
and bare current}.

We now turn to the derivation of the continuity equation,
Eq.~(\ref{eq:continuity n}) by applying Noether's theorem to the Pauli
Lagrangian coupled to the $2\times 2$-matrix Abelian gauge field $\vc{W}$.
Note that the gauge fields we are considering here are
space-time-independent Abelian class of fields satisfying
Eq.~(\ref{eq:commutator 1,2}). For this purpose, we write down the Pauli
Lagrangian for a spin one-half particle with spin-orbit coupling encoded
in a coupling to the gauge field $\vc{W}$,
\begin{equation}
  \mathcal{L}=\frac{i\hbar}{2}(\phi^{\dagger}\dot{\phi}-\dot{\phi}^{\dagger}\phi)
    - \frac{\hbar^{2}}{2m}\left(\vc{(D}\phi)^{\dagger}\cdot\vc{D}\phi\right)
    + \frac{g^{2}}{2m}\phi^{\dagger}\vc{W}\cdot\vc{W}\phi-V(\vc{x})\phi^\dagger \phi  \quad,
\end{equation}
where $\vc{D}=\vc{\nabla}-\frac{ig}{\hbar}\vc{W}$. The infinitesimal
version of the global gauge transformation, Eq.~(\ref{eq:global trans}) is
\begin{equation}
  \phi  \rightarrow   U\phi  \simeq  (1-i\vc{w}\cdot\vc{l})\phi  =  (1-i\xi \hat{n}^{a}\tau_{a})\phi
\end{equation}
where $\vc{w}$ and $\xi^{a}$ are now \emph{infinitesimal}; $\vc{\xi}$
and $\vc{\hat{n}}$ are defined by Eq.~(\ref{eq:define xi}).
Note that $\vc{\hat{n}}$ is
\emph{not arbitrary} but depends on the particular $\vc{W}$.
Since $\vc{w}$ and $\vc{W}$ still satisfy the algebra
(\ref{eq:commutator 1,2}), the above transformation is a symmetry
of the Lagrangian. The variation in the fields is
\begin{equation}
  \delta\phi=-i\xi \hat{n}^{a} \tau_{a}\phi  \quad,\qquad
  \delta\phi^{\dagger}=i\phi^{\dagger}\xi \hat{n}^{a} \tau_{a}  \quad.
\end{equation}
Noether's theorem then dictates the existence of a conserved current given by
Eq.~\eqref{eq:NoetherCurrent}. It is straightforward to show that the
associated continuity equation is
\begin{equation}
  \xi \hat{n}^{a}\partial_{t}S^{a}+\xi{\hat{n}}^{a}\partial_{i}\left(J_{i\vc{W}}^{a}\right)=0~.
\end{equation}
Since $\xi$ is arbitrary (\emph{but not} $\vc{\hat{n}}$), the continuity
equation, Eq.~(\ref{eq:continuity n}) follows.

\end{document}